\documentclass[pra,preprintnumbers,amsmath,amssymb,superscriptaddress,twocolumn]{revtex4}
\usepackage{graphicx}
\usepackage{dcolumn}
\usepackage{bm}
\usepackage{color}
\usepackage[latin9]{inputenc}
\usepackage{amsmath}
\usepackage{multirow}
\usepackage{color}

\newcommand{\ket}[2][]{\mathinner{\lvert#2\rangle}_{\hspace{-0.1em}#1}}

\begin{document}
\title{Cold-atom based implementation of the quantum Rabi model}
\author{P. Schneeweiss}
\email{schneeweiss@ati.ac.at}
\author{A. Dareau}
\affiliation{%
 \mbox{Vienna Center for Quantum Science and Technology, Atominstitut, TU Wien, Stadionallee 2, 1020 Vienna, Austria}
}
\author{C. Sayrin}
\affiliation{%
 \mbox{Vienna Center for Quantum Science and Technology, Atominstitut, TU Wien, Stadionallee 2, 1020 Vienna, Austria}
}
\affiliation{%
 \mbox{Laboratoire Kastler Brossel, Coll\`{e}ge de France, CNRS, ENS-PSL Research University, UPMC-Sorbonne Universit\'{e}s} \newline \mbox{11 place Marcelin Berthelot, 75231 Paris Cedex 05, France}}

\date{\today}
\begin{abstract}
The interaction of a two-level system (TLS) with a single bosonic mode is one of the most fundamental processes in quantum optics. Microscopically, it is described by the quantum Rabi model (QRM). Here, we propose an implementation of this model based on single trapped cold atoms. The TLS is implemented using atomic Zeeman states, while the atom's vibrational states in the trap represent the bosonic mode. The coupling is mediated by a suitable fictitious magnetic field pattern. We show that all important system parameters, i.e., the emitter--field detuning and the coupling strength of the emitter to the mode, can be tuned over a wide range. Remarkably, assuming realistic experimental conditions, our approach allows one to explore the regimes of ultra-strong coupling, deep strong coupling, and dispersive deep strong coupling. The states of the bosonic mode and the TLS can be prepared and read out using standard cold-atom techniques. Moreover, we show that our scheme enables the implementation of important generalizations, namely, the driven QRM, the QRM with quadratic coupling as well as the case of many TLSs coupled to one mode (Dicke model). The proposed cold-atom based implementation will facilitate experimental studies of a series of phenomena predicted for the QRM in extreme, so far unexplored physical regimes.
\end{abstract}
\maketitle


\begin{figure*}
	\includegraphics[]{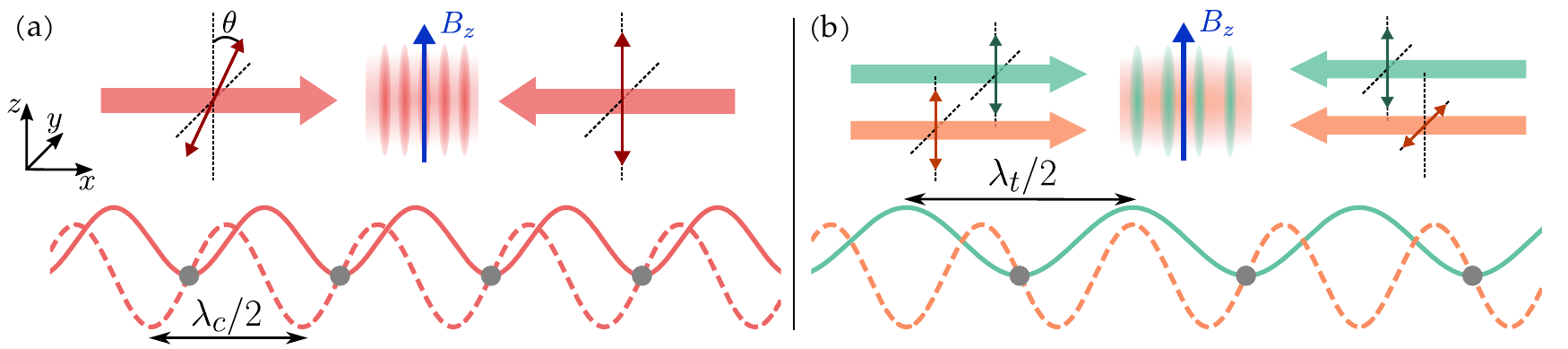}
	\caption{Sketch of the experimental implementation for the two configurations discussed in the main text. The propagation directions of the light fields are indicated by single-sided arrows, while their (linear) polarizations are indicated by double-sided arrows. The blue arrow indicates the orientation of the external homogeneous magnetic field. The trapping potential is plotted with a solid line, while the amplitude of the $x$-component of the fictitious magnetic field of the coupling lattice is plotted with a dashed line. The atoms' positions are marked with gray dots. Each trapping site realizes an implementation of the QRM Hamiltonian. (a) Lin-$\theta$-lin lattice, i.e., both lattices have the same period, $\lambda_c = \lambda_t$. (b) Configuration with two independent lattices, here for  $\lambda_t = (3/2)\lambda_c$.}
	\label{fig:exp_setup}
\end{figure*}

A two-level system (TLS) interacting with a single bosonic mode can be described at the microscopic level using the quantum Rabi model (QRM)~\cite{Rabi36,Rabi37}. Among the most well-known systems described by the QRM are single real or artificial atoms coupled to a mode of a resonator as well as single trapped ions. In the former case, the bosonic mode corresponds to microwave or optical photons while the TLS is realized by internal states of the atom. In the latter case, the quantized motion of the ion in the trap represents the bosonic mode. When the coupling strength is small enough, the rotating-wave approximation (RWA) can safely be applied and the Jaynes-Cummings (JC) model is obtained, which, arguably, is one of the most successful theoretical frameworks in quantum optics~\cite{Haroche13,Wineland13}.

More recently, there has been a growing interest in the full QRM, which is valid for arbitrary ratios of the coupling strength, $g$, and the mode frequency, $\omega$. The Hamiltonian reads
\begin{align}
\hat H &= \hbar \omega \hat a ^\dagger \hat a + \hbar g (\hat a + \hat a^\dagger)(\hat \sigma^+ + \hat \sigma^-) + \frac{\hbar \omega_0}{2} \hat \sigma_z~,
\label{eq:QRM}
\end{align}
with the bosonic creation (annihilation) operators $\hat a^\dagger$ ($\hat a$), the TLS's raising (lowering) operators $\hat \sigma^+$ ($\hat \sigma^-$), the Pauli matrix $\hat \sigma_z$, and the energy of the TLS $\omega_0$. Despite its fundamental nature, an analytic solution for the spectrum of the QRM was only found recently~\cite{Braak11}. Remarkably, for a large enough coupling strength, qualitatively new phenomena~\cite{Casanova10} such as the excitation of two atoms with one photon~\cite{Garziano16} are predicted and novel protocols for quantum information processing and quantum communication have been proposed~\cite{Nataf11,Romero12,Kyaw15}. A quantum phase transition that is predicted to occur in the regime of large dispersive coupling attracted special interest, too~\cite{Hwang15}.

The QRM in the regime where $g$ is a significant fraction of $\omega$ (ultra-strong coupling, USC) has been entered with several experimental systems including quantum wells~\cite{Anappara09,Guenter09,Todorov10,Zhang16}, superconducting systems~\cite{Bourassa09,Niemczyk10,Forn-Diaz10}, and molecular ensembles in cavities~\cite{Schwartz11,George16}. Very recently, even ratios $g/\omega>1$ have been reached~\cite{Yoshihara17,Forn-Diaz17}, which corresponds to the deep-strong coupling (DSC) regime. Note that, strictly speaking, the classification of coupling regimes also depends on the number of excitations in the system~\cite{ArXiv_Rossatto16}. Despite this tremendous experimental progress, it is an open challenge to find fully versatile experimental implementations of the QRM. Ideally, these allow for a widely tunable dynamical adjustment of the model parameters as well as provide ways for the preparation and read-out of the quantum state of the system. In this context, a number of dedicated simulators have been studied, including approaches where suitably manipulated ensembles of cold atoms mimic the QRM dynamics~\cite{Felicetti17}.

Here, we describe a way to implement the QRM using single trapped cold atoms exposed to a suitable magnetic field pattern. Our approach allows one to dynamically tune the system parameters, $g$, $\omega$, and $\omega_0$, relative to each other over a wide range. Remarkably, assuming realistic experimental conditions, our cold-atom based implementation of the QRM also enables access to the regimes of ultra-strong and deep strong coupling as well as dispersive DSC, the latter requiring that $\omega_0 \gg \omega$ while $g>\omega$~\cite{Felicetti17}. The initialization and read-out of the TLS's and the bosonic mode's states can be achieved by means of established cold-atom techniques~\cite{Cohen-Tannoudji11}. We quantitatively discuss an example realization based on individual, cold Rubidium atoms confined in a one-dimensional optical lattice. Finally, we discuss the implementation of important generalizations of the QRM.

In order to introduce the underlying principle of our approach, we initially consider a canonical TLS, a spin-1/2 particle, confined in a 1D harmonic trap of frequency $\omega$ and exposed to a tailored magnetic field. The Hamiltonian reads
\begin{equation}
\hat H = \hbar \omega \hat a ^\dagger \hat a + g_L \mu_B \bm{B} \cdot \hat{\bm{S}}/\hbar~,\label{eq:Zeeman} 
\end{equation}
with $\hat{\bm{S}}= (\hbar/2) \hat{\bm{\sigma}}$ and $\hat{\bm{\sigma}}=(\hat \sigma_x,\hat \sigma_y,\hat \sigma_z)$ the Pauli matrices, $\mu_B$ the Bohr magneton, $g_L$ the Land\'{e} factor, and $\bm{B}$ a position-dependent magnetic field. We chose $\bm{B} = (b_x \hat x, 0, B_z)$, i.e., the magnetic field has a constant component along $z$ and a $x$-component that varies linearly along the $x$-coordinate with a gradient $b_x$. While such a pattern cannot be implemented with real magnetic fields only ($\bm{\nabla} \cdot \bm{B} = b_x \neq 0$), it can be realized by combining real and so-called fictitious magnetic fields, originating from the vector ac-Stark shift~\cite{Cohen-Tannoudji72,Deutsch10}. Fictitious magnetic fields can be obtained when a multi-level atom is exposed to a detuned light field. The induced magnetic field is maximal for circular polarization of the light and vanishes for linear polarization. In particular, appropriate fictitious field patterns can be generated with certain optical lattice configurations~\cite{Hamann98,Mandel03} and are naturally present in strongly confining optical dipole traps~\cite{Kaufman12,Thompson13,Albrecht16}. With this magnetic field pattern, the Hamiltonian becomes
\begin{align}
\hat H &= \hbar \omega \hat a ^\dagger \hat a + g_L \mu_B/2 \; (b_x \hat x \hat \sigma_x + B_z \hat \sigma_z)~.
\label{eq:mag}
\end{align}
We can rewrite (\ref{eq:mag}) using $\hat x=x_0 (\hat a + \hat a ^\dagger)$ with \mbox{$x_0=\sqrt{\hbar/(2M\omega)}$}, where $M$ is the particle's mass. Moreover, we use $\hat \sigma_x =\hat \sigma^+ +\hat \sigma^-$ and then obtain precisely the Hamiltonian of the QRM given in~(\ref{eq:QRM}). In this approach, the bosonic mode corresponds to motional states of the atom in the harmonic potential of frequency $\omega$. The coupling strength and energy of the TLS can both be adjusted and are given by $g=(\mu_B g_L b_x x_0)/(2\hbar)$ and $\omega_0=\mu_B g_L B_z/\hbar$, respectively.


In the following, we quantitatively discuss one out of several possible implementations of the Hamiltonian~\eqref{eq:mag}. We consider multi-level atoms of spin $F$ that are in their electronic ground state. For example, this could be alkali atoms which are commonly used in cold-atom experiments. We assume an optical lattice resulting from the interference of two counter-propagating laser beams that are characterized by their wave number $k_t=2\pi/\lambda_t$ and which are linearly polarized along the same axis, see Fig.~\ref{fig:exp_setup}. The induced trapping potential is proportional to the intensity of the resulting standing wave via the atom's scalar polarizability~\cite{Deutsch10}. We refer to this lattice as the \emph{trapping lattice}. We assume that the trapping sites are loaded in such a way that each lattice site is occupied by, at most, one atom~\cite{Schlosser02}. In order to induce a coupling between the spin and motional degrees of freedom, we consider another optical lattice, called the \emph{coupling lattice}, consisting of two counter-propagating laser beams (wave number $k_c=2\pi/\lambda_c$) with orthogonal linear polarizations. The intensity of the combined field is uniform along the $x$-direction but the polarization changes with position. In this case, the atom experiences a spatially varying vector ac-Stark shift~\cite{Mandel03}, equivalent to the Zeeman interaction with a fictitious magnetic field $\bm{B}_\mathrm{fict} (\hat x)=B_x \sin \left[ 2 k_c \hat x \right]\bm{e}_x$, with $\bm{e}_x$ the unit vector along $x$. The total Hamiltonian including the kinetic energy of the atom, the contributions of the trapping and coupling lattices, and the Zeeman shift due to an external homogeneous offset magnetic field oriented along the $z$-direction, $B_\mathrm{z}$, reads
\begin{align}
	\hat H = \frac{\hat p^2}{2M} &+ \frac{V_0}{2}\left( 1-\cos[2k_t \hat x] \right) \nonumber \\
	&+ \frac{g_F \mu_B}{2} \left( B_x \sin \left[ 2 k_c \hat x \right] \hat F_x + B_z \hat F_z \right)~,
	\label{eq:latticeHamiltonian}
\end{align}
where $V_0$ is the trap depth, $\hat F_x$ and $\hat F_z$ are spin angular momentum operators, and $g_F$ is the Land\'{e} factor of the hyperfine level. The wavelengths of the trapping and coupling lattices as well as their relative phase are chosen such that the local minima of the trapping sites coincide with the zero crossings of the fictitious magnetic field. The trapping lattice is assumed to be sufficiently deep, such that tunneling and interactions between atoms in neighboring sites can be neglected. Near the local minima, the trapping potential can well be approximated by a harmonic potential with a frequency $\omega = 2\sqrt{V_0 E_\mathrm{r}}/\hbar$ where $E_\mathrm{r} = \hbar^2 k_t^2/(2M)$ is the recoil energy associated with the trapping lattice. Around these minima, the fictitious field is well approximated by a linear gradient $\bm{B}_\mathrm{fict} \sim  b_x x \bm{e}_x$, with $b_x = 2B_xk_c$. With these approximations, we obtain an array of trapping sites, where each site realizes the QRM Hamiltonian~\eqref{eq:QRM}, whose parameters can be widely tuned: the frequency, $\omega$, can be adjusted with the depth of the trapping lattice ($\omega \propto \sqrt{V_0}$), which also involves a change of the coupling strength $g$ via $x_0(\omega)\propto V_0^{-1/4}$. The coupling strength can be adjusted independently from $\omega$ by tuning the intensity, $I_c$, of the coupling lattice ($g \propto I_c$), while the transition energy of the atom, $\omega_0$, can be adjusted via $B_z$ ($\omega_0 \propto B_z$). The parameters $\omega$, $g$, and $\omega_0$ define the regime in which the QRM is implemented. As quantitatively discussed in the following, our approach allows not only the experimental exploration of the well-known JC regime ($g\ll\omega$) but also the extreme regimes of ultra-strong coupling, deep strong coupling as well as dispersive deep strong coupling.

To be specific, we now consider the case of the Rubidium isotope $^{87}\mathrm{Rb}$ in its electronic ground state, $5S_{1/2}$. This type of atom offers particularly easy experimental handling. In order to obtain a large fictitious magnetic field, the wavelength of the coupling lattice, $\lambda_c$, can be chosen close to the Rubidium $D_1$ and $D_2$ lines. The simplest choice for the wavelength of the trapping lattice, $\lambda_t$, is then $\lambda_t = \lambda_c$. In this case, both lattices can be generated with a single pair of counter-propagating beams with linear polarizations tilted with respect to each other (``lin-$\theta$-lin'' configuration), see Fig.~\ref{fig:exp_setup}a. The relative strength of the trapping and coupling lattices, and, thus, the ratio $g/\omega$, can be tuned by varying the angle $\theta$ between the two polarizations. In the case that the trapping and the magnetic lattices are generated by two independent light fields, a matching of the local trap minima with the magnetic field zero crossings can be achieved by setting $\lambda_t = 2\lambda_c$. An interesting choice for $\lambda_c$ in this case is the so-called tune-out wavelength ($790.0\,$nm for Rb), at which the scalar polarizability of the atom vanishes and, in which case, the coupling lattice only induces a fictitious magnetic field. Yet another choice of wavelength of the trapping lattice is $\lambda_t = (3/2)\lambda_c$, which also realizes the Hamiltonian~\eqref{eq:QRM} on each site, see Fig.~\ref{fig:exp_setup}b. In this case, the sign of $g$ changes with every lattice site. 
\begin{table}
		\normalsize
		\vspace{0.5cm}
		\begin{tabular}{|c||c|c|c|}
			\hline
			Parameter & (a) lin-$\theta$-lin & (b) two-lattices & unit\\
			\hline
			$\lambda_t$ & \multirow{2}{*}{787} & 1185 & nm \\ 
			$\lambda_c$ & &  790.04 & nm \\
			\hline
			specific & $P = 2.6\,\mathrm{W}$ & $P_t = 14\,\mathrm{W}$ & - \\
			configuration & $\theta \approx 49^\circ$ & $P_c = 0.75\,\mathrm{W}$ & - \\
			\hline
			$V_0$ & 1$\times 10^5$ & 2$\times 10^5$ & $E_\mathrm{r}$ \\
			$\omega_\mathrm{eff}/(2\pi)$ & 2.9 & 2.2 & $\mathrm{MHz}$\\
			$g_\mathrm{eff}/(2\pi)$ & 8.5 & 6.5 & $\mathrm{MHz}$\\
			\hline
			scattering &  \multirow{2}{*}{$\Gamma = 16.9$} & $\Gamma_\mathrm{t}=0.09$ & $\mathrm{kHz}$\\
			rate &  & $\Gamma_\mathrm{c} = 3.6$ & $\mathrm{kHz}$\\
			\hline
		\end{tabular}
		\caption{Parameters for the two proposed experimental configurations based on $^{87}$Rb. In both cases, we consider a waist of $w=15\,\mathrm{\mu m}$ for the laser beams generating the lattices. In the lin-$\theta$-lin configuration (a) we indicate the power $P$ per laser beam and the relative angle $\theta$ between the linear polarizations of the two beams. For configuration (b), $P_t$ is the power per laser beam used to form the trapping lattice, and $P_c$ is the respective power for the coupling lattice. The parameters $\omega_\mathrm{eff}$ and $g_\mathrm{eff}$ are, respectively, the effective bosonic mode frequency and coupling strength (see main text). The rates $\Gamma_\mathrm{t}$ and $\Gamma_\mathrm{c}$ ($\Gamma=\Gamma_\mathrm{t}+\Gamma_\mathrm{c}$) quantify inelastic scattering induced by the trapping and the coupling lattice, respectively.} 
		\label{tab:parameters}
\end{table} 

A set of example parameters calculated for $^{87}\mathrm{Rb}$ and based on the two discussed configurations is presented in Tab.~\ref{tab:parameters}. Remarkably, for laser field configurations accessible with current technology, we obtain ratios $g/\omega\approx 3$, i.e., clearly in the deep strong coupling regime. Our scheme allows a fully flexible choice of $\omega_0$ by adapting the homogeneous $B_z$-component of the magnetic field. For instance, in the case of configuration (a) in Tab.~\ref{tab:parameters}, resonance, $\omega=\omega_0$, is obtained for $B_z \approx 4\,$G. The lattice light fields can be inelastically scattered by the atoms with rate $\Gamma_\mathrm{t}$ in the case of the trapping lattice and with $\Gamma_\mathrm{c} \propto I_c$ for the coupling lattice. The ratio $g/\Gamma_\mathrm{c}$ is a constant. We estimate that both rates, $\Gamma_\mathrm{t}$ and $\Gamma_\mathrm{c}$, are orders of magnitude smaller than the coupling rate $g$ for our settings. For this reason, inelastic scattering is negligible during many cycles of coherent evolution of the system. Other sources of decoherence include heating of the atoms in the trapping potential as well as magnetic field noise. In typical optical-lattice setups, both effects are irrelevant on the time scales considered here.

As the trapping potential is necessarily finite, deviations from the harmonic approximation are expected for large enough energies. In particular, the atomic center-of-mass wavefunctions and energies of highly excited motional states will deviate from those of the harmonic oscillator. Moreover, it is clear from Fig.~\ref{fig:exp_setup} that the fictitious magnetic field is well approximated by a linear gradient only close to the field zero crossings. In order to quantify when these finite size effects become important, we perform a numerical diagonalization of the lattice Hamiltonian (\ref{eq:latticeHamiltonian}) in position-spin space~\cite{Johansson12}, and compare the obtained eigenenergies and eigenfunctions with the ones of the QRM, i.e., for an ideal harmonic oscillator trapping potential and a linear fictitious field gradient. For this comparison, we use effective values, $g_\mathrm{eff}$ and $\omega_\mathrm{eff}$, obtained as follows. When setting $B_z=0$, the Hamiltonian~\eqref{eq:latticeHamiltonian} is diagonal in the $\hat{F}_x$ basis. We then fit the effective potential for the high-field seeking Zeeman sub-state near its local minimum. The local curvature then determines $\omega_\mathrm{eff}$ while the position of the minimum yields $g_\mathrm{eff}$. The discrepancy between the theoretical and effective $g/\omega$ ratio increases for larger values of the coupling strength. For the extreme configurations presented in Tab.~\ref{tab:parameters}, this discrepancy is about $10\,\%$.

\begin{figure*}
	\includegraphics[width=\textwidth]{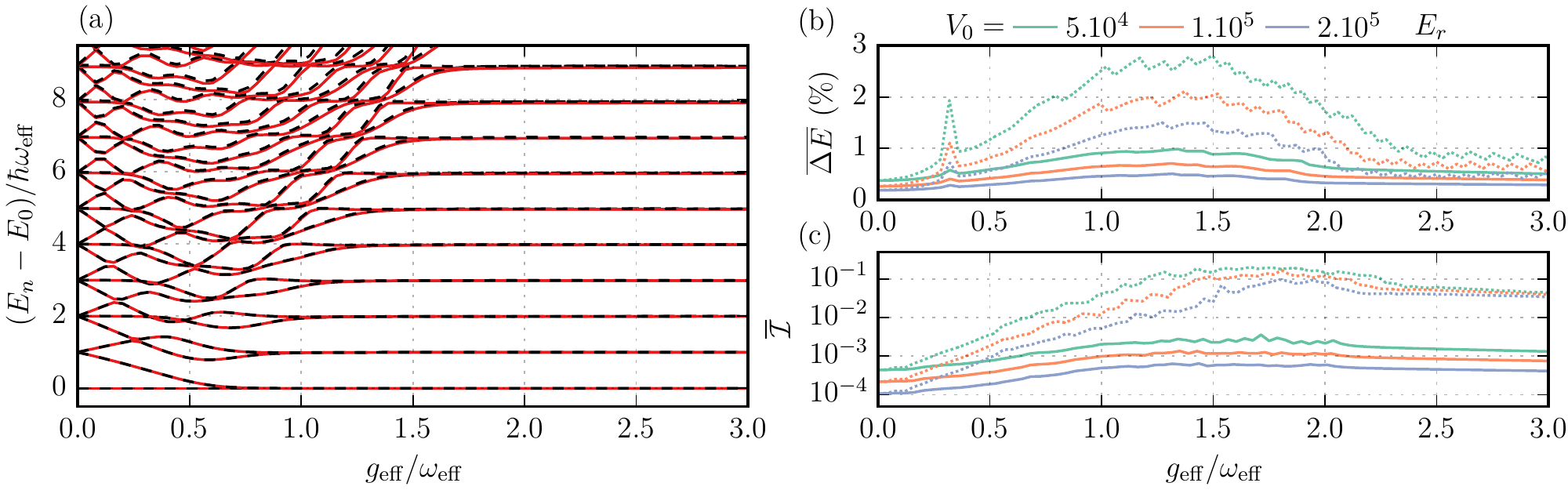}
\caption{Comparison between the ideal QRM in the position-spin representation and our experimental implementation. (a): Energy spectrum of the 30 first states as a function of the coupling strength $g_\mathrm{eff} / \omega_\mathrm{eff}$, for the QRM (black dashed lines) and the ``lin-$\theta$-lin'' lattice implementation (solid red lines). Experimental parameters correspond to configuration (a) in Tab.~\ref{tab:parameters}. (b) and (c): Comparison of the eigenenergies (b) and eigenfunctions (c) of the 30 first states of the ideal QRM model and the two discussed experimental implementation: ``lin-$\theta$-lin'' (solid lines) and ``two lattices'' (dashed lines). Different colors correspond to different depths $V_0$ of the trapping lattice. For each set of parameters, we numerically compute the eigensolutions for the QRM $\left\{E_n^\mathrm{th}, |\psi_n^\mathrm{th}\rangle \right \}$ and the experimental implementation $\left\{E_n^\mathrm{exp}, |\psi_n^\mathrm{exp}\rangle \right \}$. We then compute the mean relative energy discrepancy $\overline{\Delta E} = (1/N) \sum_{n<N} |1-E_n^\mathrm{exp}/E_n^\mathrm{th}|$ (b) and the mean state infidelity $\mathcal{\bar I}= (1/N) \sum_{n<N} 1 - |\langle\psi_n^\mathrm{th}|\psi_n^\mathrm{exp}\rangle|^2$ (c). We see excellent agreement in the considered sub-space. All calculations are done on resonance, $\omega=\omega_0$.}
	\label{fig:simulation}
\end{figure*}

The results of a systematic comparison between the QRM and our experimental implementation are summarized in Fig.~\ref{fig:simulation}, accounting for several trapping lattice depths and for $0 \leq g_\mathrm{eff} / \omega_\mathrm{eff}\leq 3$. For every configuration, we compare the first 30 eigenstates, corresponding to the first 10 motional states in the case of $^{87}$Rb in the $F=1$ hyperfine state. For the parameters of Tab.~\ref{tab:parameters}, the mean agreement of the eigenenergies is better than $1\,\%$  and $2\,\%$, respectively, for the ``lin-$\theta$-lin'' and the ``two-lattices'' configuration over the full considered range of $g_\mathrm{eff} / \omega_\mathrm{eff}$. The mean infidelity of the eigenfunctions is less than $2\cdot 10^{-3}$ and $10^{-1}$, respectively. The results are essentially unchanged when we vary $\omega_0$, enabling also the study of large dispersive coupling.

A sequence for a simple experimental test of the system would be to first start inducing the familiar Rabi oscillations encountered in the JC regime. For this purpose, a single atom is initialized in the motional ground state of the harmonic trap using standard techniques~\cite{Cohen-Tannoudji11}. After the cooling is completed, the atom is optically pumped into an energetically higher-lying Zeeman sub-state. In order to start the Rabi oscillations, the coupling $g$ is switched on abruptly by rapidly ramping up the coupling lattice. Then, the atomic population oscillates between the different internal states until the coupling is switched off again. The state population can then be measured by, e.g., state-selective optical read-out. A full tomography of the internal state of the atom can be performed, for example, using stimulated Raman adiabatic passage techniques~\cite{Vewinger03,Volz06}. The occupation of the bosonic mode can be obtained by determining the population of the motional states of the trap~\cite{Morinaga99,Kaufman12,Belmechri13,Thompson13}. The experiment can then be repeated with larger and larger ratios of $g/\omega$ and different detunings, allowing the experimental study of genuine QRM effects~\cite{Casanova10,Wolf13}. Although there has been impressive progress on the experimental study of the QRM, most experiments are, so far, limited to spectroscopic analyses. Dynamics signatures have, by now, been measured for $g/\omega \approx 0.1$~\cite{Zhang16}, and the dynamics for larger coupling strength has been observed in a digital quantum simulation~\cite{ArXiv_Langford16}. Our approach gives direct access to the QRM dynamics and should, for example, facilitate the direct observation of the collapse and revival predicted in the DSC regime~\cite{Casanova10}. Another option enabled by our in-situ control of the system parameters is the adiabatic preparation of the ground state of the QRM, which exhibits entanglement between the bosonic mode and the TLS, given a large enough coupling strength~\cite{ArXiv_Rossatto16}. These examples emphasize the new opportunities created by our approach.


Besides the implementation of the QRM Hamiltonian~(\ref{eq:QRM}), our scheme can be adapted to implement important variations of the QRM. For example, in addition to the homogeneous magnetic field along the $z$-direction, one can introduce a constant component along $x$ of strength $\epsilon$, c.f. Fig.~\ref{fig:exp_setup}. The combined pattern composed of external and fictitious magnetic fields then reads $\bm{B} = (b_x \hat x + \epsilon, 0, B_z)$. With that, we obtain the Hamiltonian given in (\ref{eq:latticeHamiltonian}) as well as an additional term $g_\epsilon \hat F_x$ with $g_\epsilon=\epsilon \mu_B g_F /2$. This resembles the so-called generalized or driven Rabi model. One of many interesting phenomena predicted for this system is the emergence of Dirac cone-like intersections in the system's energy landscape, which are expected to exhibit a non-zero geometric phase~\cite{Batchelor16}. Note that this setting reduces to the well-known state-dependent optical lattice~\cite{Mandel03} for $B_z=0$.

So far, we overlapped the zero-crossings of the coupling lattice with the minima of the trapping lattice. Another experimental option is to spatially match the local extrema of both lattices by adapting their relative phase. Then, the atoms are exposed to a fictitious magnetic field with a curvature, $B_x=b_{xx} \hat x^2$. This gives rise to a coupling $\hbar g_2 (\hat a + \hat a^\dagger)^2 \hat F_x$ with $g_2=\mu_B g_F b_{xx} x_0^2/(2\hbar)$. For quadratic coupling, the emergence of dark-like states~\cite{Peng17} has been predicted recently. Moreover, a spectral collapse~\cite{Ng99,Emary02}, for which all eigenenergies of the system approach a common value, is among the most remarkable effects of this model. We expect that our approach will allow the experimental study of quadratic coupling. The tunneling between neighboring lattice sites, however, might have to be taken into account for large $g_2$ as it leads to a reduction of the effective trap frequency and might reduce the height of the energy barrier between two sites.

Until now, we have considered an experimental implementation using $^{87}\mathrm{Rb}$, whose lower hyperfine ground state has a spin of $F=1$ while the QRM considers a TLS. A spin of precisely $1/2$ is, for example, encountered for $^6\mathrm{Li}$ in the lower hyperfine ground state. Lithium is commonly used in cold-atom experiments, and important techniques such as ground-state cooling have been demonstrated in optical lattices~\cite{Omran15}. However, heavy alkali atoms offer a few practical advantages such as easier laser cooling and imaging. Moreover, they also feature large fine-structure splittings which offer a more favorable ratio $|\bm{B}_\mathrm{fict}|/\Gamma_\mathrm{c}$ when implementing fictitious magnetic fields via a tune-out light field. Different means have been developed to constrain a system to a sub-Hilbert space, realizing a so-called quantum Zeno dynamics~\cite{Raimond10}. For example, using Raman coupling between the two hyperfine manifolds, the coherent evolution of Rabi oscillations between the five Zeeman sub-states in $^{87}$Rb, $F=2$ has been restricted to $\ket{F=2,m_F=1}$ and $\ket{F=2,m_F=2}$ only, effectively realizing a spin-$1/2$ system~\cite{Schaefer14}. In this way, the QRM can be implemented while benefiting from the advantages of heavier alkalis.

Working with atoms of higher-dimensional spin is of scientific interest on its own as it enables the experimental study of the Dicke model. This model considers $N$ identical spin-1/2 particles coupled to a common bosonic mode and is valid for arbitrary ratios $g/\omega$ as no RWA is applied. The Hamiltonian reads
\begin{align}
\frac{\hat H_{D}}{\hbar} &= \omega \hat a ^\dagger \hat a + \frac{\omega_0}{2} \sum_{i=1}^N \hat \sigma_{i,z} + \frac{g}{\sqrt{N}} (\hat a + \hat a^\dagger)\sum_{i=1}^N \hat \sigma_{i,x}~.
\label{eq:Dicke}
\end{align}
We can introduce $\hat F_{x}=1/2\sum_i \hat \sigma_{i,x}$ and $\hat F_{z}=1/2\sum_i \hat \sigma_{i,z}$, which are angular momentum operators of a higher-dimensional spin-$N/2$. As~(\ref{eq:Dicke}) applied on an angular momentum eigenstate $\ket{F,m_F}$ does not change the quantum number $F$, a single spin-$F$ particle equivalently represents the Dicke model with $N=2F$ particles in the sub-space spanned by the states $\ket{F,m_F=-F \ldots F}$. In this sense, the isotope $^{85}$Rb in the hyperfine state $F=3$ would allow one to simulate the Dicke model for $N=6$. One of the important phenomena predicted for the Dicke model is a quantum phase transition that is expected to occur at large enough coupling strength~\cite{Emary03b}. It has been shown that signatures of this effect prevail for a finite-size system constrained to the largest pseudo-spin sub-space~\cite{Emary03b}, such that it might be observable with our approach.


In summary, we have proposed a cold-atom based platform for the experimental investigation of the QRM including its dynamics. Remarkably, assuming realistic experimental conditions, our estimations predict that the implementation of the QRM in the regimes of ultra-strong, deep strong and dispersive deep strong coupling should be feasible. Corresponding experiments can take advantage of the rich toolbox developed in cold-atom physics, facilitating, e.g., simple state preparation and read-out of the system. Moreover, we have presented ways to implement important generalizations of the model.

Future theory work might conceive extensions of our scheme to further generalizations. For example, effective spin-spin interactions in the Dicke model~\cite{Jaako16} might be introduced by applying an additional light field that gives rise to a tensorial ac-Stark shift of the atomic levels~\cite{Deutsch10}. This should then yield the intended coupling $\propto F_x^2$. Moreover, our approach should allow the ultra-strong coupling to two bosonic modes, which might, e.g., enable studying the Jahn-Teller instability with cold atoms~\cite{Larson08,Meaney10}. Finally, the QRM in the presence of dissipation exhibits surprising, non-trivial effects~\cite{Beaudoin11}. Our approach might open up novel ways to their experimental study and provide means to develop tools for quantum reservoir engineering~\cite{Poyatos96} in the USC and DSC regime.

We thank C.~Clausen, I.~Mazets, P. Rabl, A. Rauschenbeutel, M.~Sanz, E.~Solano, and J.~Volz for stimulating discussions and helpful comments. Financial support by the European Research Council (Consolidator Grant NanoQuaNt and Marie Curie IEF Grant
328545) is gratefully acknowledged.

\bibliography{ATI_Schneeweiss}
\end{document}